\documentclass[10pt,journal,compsoc]{IEEEtran}
\pdfoutput=1
\usepackage{verbatim}
\usepackage{verbatim}
\usepackage{graphicx}
\newcommand{\norm}[1]{\left\lVert#1\right\rVert}

\newtheorem{theorem}{Theorem}

\newtheorem{lemma}[theorem]{Lemma}

\newtheorem{remark}{Remark}
\usepackage{amssymb}
\usepackage{dsfont}
\usepackage{multirow}
\usepackage{color}
\usepackage{multicol}
\usepackage{mathtools}
\usepackage[mathscr]{euscript}

\usepackage{cite}
\usepackage{hyperref}

\usepackage{breakurl}

\usepackage[tight,footnotesize]{subfigure}

\ifCLASSINFOpdf
\else
\fi
\usepackage{amsmath}

\begin{document}
\title{Will Scale-free Popularity Develop Scale-free Geo-social Networks?}

\author{
  Dong Liu,
  Viktoria Fodor,~\IEEEmembership{Member, IEEE}
  and~Lars~K. Rasmussen, \IEEEmembership{Senior Member, IEEE}
  \IEEEcompsocitemizethanks{
    \IEEEcompsocthanksitem The authors are with School of Electrical Engineering and Computer Science, Access Linnaeus Centre,
    KTH Royal Institute of Technology, Stockholm, SE-100~44,
    Sweden. E-mail: \{doli, vfodor, lkra\}@kth.se}}

\IEEEtitleabstractindextext{%
  \begin{abstract}\label{abstract}
    Empirical results show that spatial factors such as distance,
    population density and communication range affect our social
    activities, also reflected by the development of ties in social
    networks. This motivates the need for social network models that take
    these spatial factors into account. Therefore, in this paper we
    propose a gravity-low-based geo-social network model, where
    connections develop according to the popularity of the individuals,
    but are constrained through their geographic distance and the
    surrounding population density. Specifically, we consider a power-law
    distributed popularity, and random node positions governed by a
    Poisson point process. We evaluate the characteristics of the emerging
    networks, considering the degree distribution, the average degree of
    neighbors and the local clustering coefficient. These local metrics
    reflect the robustness of the network, the information dissemination speed
    and the communication locality.
    We show that unless the communication range is strictly limited,
    the emerging networks are scale-free, with a rank exponent
    affected by the spatial factors. Even the average neighbor degree and
    the local clustering coefficient show tendencies known in
    non-geographic scale-free networks, at least when considering
    individuals with low popularity. At high-popularity values, however, the spatial constraints lead to popularity-independent average neighbor degrees and clustering coefficients.
  \end{abstract}

  \begin{IEEEkeywords}
    Social tie, network model, popularity, geographic distance, degree
    distribution, clustering coefficient.
  \end{IEEEkeywords}
}


\maketitle


%
\IEEEpeerreviewmaketitle

\IEEEraisesectionheading{\section{Introduction}\label{sect_intro}}
\IEEEPARstart{S}{ince} the first empirical study in $1967$ by Jeffrey Travers and Stanley Milgram
\cite{travers1967small} about forwarding letters over social ties, the structure of complex networks
has been investigated extensively \cite{watts2004six}. Since a significant part of network traffic is now generated by social network applications, the characterization of these social networks is becoming increasingly important, both for the design and dimensioning of future networks, and to estimate the information spreading capability of the emerging network structures.

With the fast growing Internet and cellular networks, the emergence of
social networks and a
variety of mobile applications, our communication patterns are now well recorded and analysed.
For example, the small world phenomenon has been investigated in large scale compared with Stanley Milgram's experiment, in the Microsoft Messenger network \cite{leskovec2008planetary}, and in the Facebook social network \cite{backstrom2012four}, showing an average distance, or path length between pairs of individuals of $6.6$ and $4.74$ respectively.

While networks with short path lengths can emerge in various ways, in
social networks the main reason seems to be the scale-free
distribution of node degrees \cite{ebel2002scale,catanese2012,muchnik2013,laniado2011co}.
The emergence of
the scale-free distribution has been motivated by the preferential
attachment in growing networks, in the seminal work of Barabasi and
Albert \cite{barabasi1999emergence}, but also through considering inherent
characteristics of the individuals, that affect their
\emph{popularity} in the social network
\cite{caldarelli2002scale,boguna2003class,muchnik2013,laniado2011co}.

In addition to popularity, connectivity patterns are shown to be affected by the physical locations of the individuals, due to decreased interest in connecting to very distant peers \cite{liben2005geographic,scellato2011socio,deville2016scaling}, or in the case of infrastructure-less communication, due to the limitations of the wireless networking technology \cite{6847726,6839046,li2014social,costa2008socially}.

%
%
%
%

In this paper we derive analytic models of network structures that
emerge when these spatial effects are taken into account. We consider
a scale-free popularity distribution \cite{catanese2012,muchnik2013,laniado2011co}
and mutual-interest based social ties \cite{caldarelli2002scale,boguna2003class, muchnik2013},  however, modified through the effect of the physical distance of the individuals, and
the density of the population \cite{scellato2011socio}.  Specifically, we consider two kinds of effects: first, the decreased willingness of individuals to form social tie over increasing distances, and second, the physical limitations of direct communication in wireless networks.

We characterize the emerging network through the degree distribution, the average neighbor degree, and the clustering coefficient. Based on analytical and numerical results we conclude that the node locations can have significant influence on the emerging network structure.
We show, that the degree distribution remains scale-free, unless the
communication range of the individuals is strongly limited. In this case, a Poisson degree distribution emerges.

The average neighbor degree first decreases with increasing node popularity, then remains constant due to the spatial limitations.
Looking at the neighbor degree as a function of the node degree, the emerging networks are disassortative, unless the geographic limitations are very strict. The local clustering coefficient shows similar tendency. In summary, we find that spatial effects do not alter the main characteristics of social networks, unless the communication distance is strongly limited. In that case the emerging networks resemble random lattice networks.

%
%



The remaining part of this paper is organized as follows. We present
the relevant research and work in Section~\ref{sect_relatedWork}, and introduce the considered metrics in Section \ref{sec:metrics}.
In Section~\ref{sec:systemModel}, we explain how the individuals are modeled
by a marked Poisson process and what criteria we use to develop social
ties, which is followed by theoretical analyses in
Section~\ref{modelAnalyses}. We present numerical results in
Section~\ref{sec:num_results}. In Section~\ref{sec:results_analysis} we discuss how the analytic results
are reflected in large scale measurements, and conclude the paper in Section~\ref{sec:conclusion}.

\section{Related Work}\label{sect_relatedWork}

The first attempt to describe communities with random connection
patterns dates back to the Erdos-Renyi (ER) random graph model from
1959 \cite{erdos1960}. This model, however, failed to describe some
inherent properties of networks that emerged through social
connections: the ER model shows the desired small-world phenomenon, but
not the clustering experienced in real networks. This called for more
complex network structures, such as the Watts-Strogatz (WS) small-world
network \cite{watts1998}, providing high clustering and tunable
average path length, still preserving homogeneous connectivity with a
Poisson degree distribution. It has been recognized, though that many
of the networks, from biological structures to the Internet, are not
homogeneous, but scale-free \cite{barabasi2016network}, where the degree distribution follows a
power law $k^{-\gamma}$, $k$ denoting the degree and $\gamma$ the
scaling exponent, with typical values of $2 < \gamma <
3$.

The most well-known example of constructing a scale-free topology is the Barabasi-Albert model \cite{barabasi1999emergence},
where the network is formed by adding new vertices, and connect them through a preferential attachment strategy.
The link preference in the BA model is proportional to the degree of vertices already in network, which results in a
scale-free network with scaling exponent $\gamma=3$. More flexible versions of the preferential attachment model are presented in \cite{dorogovtsev2000structure}\cite{newman2001random}\cite{buckley2004popularity}.


Another hypothesis of the emergence of scale-free networks is based on the inherent \emph{popularity} of the members of the network. This hypothesis has been proposed in \cite{caldarelli2002scale,boguna2003class} and is extensively tested for interactions of Wikipedia contributors in \cite{perra2012activity,muchnik2013,laniado2011co}, showing that the activity level of the contributors deterministically affects the mean degree, that is, the mean number of neighbors in the interaction graph.

Neither the preferential attachment, nor the popularity based models
can directly capture the effect of geographic distance on the structure of social networks. Several studies show, however, that the probability of a social connection depends
on the geographic distance of the individuals. Statistics on bloggers of
Live Journal in USA \cite{liben2005geographic} show the probability $P(d)$
of social connection of individuals in a geo-distance of $d$ is similar to
$d^{-1}+\epsilon$, phone call statistics of a large mobile networks show
$P(d) \sim d^{-\varrho}$, $\varrho=1.3-2$ \cite{lambiotte2008geographical,deville2016scaling}. Similarly,
investigations of social network ties indicate $\varrho=0.5-1.03$ \cite{backstrom2010find,goldenberg2009distance,scellato2011socio}.
That is, the probability of social ties decreases with distance, and is more constrained
in mobile than in online social networks.

The detailed studies in \cite{liben2005geographic,scellato2011socio,deville2016scaling} reveal additional factors that need to be considered for a geo-social network model, showing, that both the network density and the node popularity tune the influence of the distance. Individuals tend to establish longer distance links if they have only few nearby neighbors, and individuals with many friends are more likely to connect over long distances.
To combine all these parameters,
\cite{krings2009urban,scellato2011socio,barthelemy2011spatial} propose
to follow a \emph{gravity law} like attachment rule, where, like in the case of Newton's gravity law, the attraction of two individuals is proportional to the product of their popularity, and inversely proportional to the square of their distance.

Closest to our work in modeling geo-social networks is \cite{flaxman2004}, where BA type preferential attachment is combined with geometric limitations, considering a deterministic maximum connection radius, that is, addressing the case when the networking technology limits the possibility of building up a connection. Consequently, \cite{flaxman2004} addresses one of our concerns, but can not reflect all the characteristics of geo-social graphs, where distance affects the willingness of the individuals to form social ties.

\section{Metrics of Interest}\label{sec:metrics}

In this paper we characterise the geo-social network, where social ties emerge
based on popularity, geographical distance and population density. We evaluate the
\emph{degree distribution} of the nodes, to see whether the scale-free property, often found in social networks is retained.

We investigate
the \emph{average neighbour degree} of nodes with given popularity and degree, to evaluate, whether the emerging network is assortative, that is, the degree of the neighbors is positively correlated with the node's own
degree, disassortative, showing negative correlation, or neutral.

Assortative networks has been found to be resilient to the removal of high degree nodes, presumably because high degree nodes are clustered together anyway. Due to the same reason, epidemics are preserved better in assortative networks, while the spreading of the epidemics is smaller than in disassortative or neutral ones \cite{newman2002assortative}.
ER random graphs and large BA scale-free graphs, constructed through preferential attachment have been shown to be neutral \cite{barrat2005rate}, while both positive and negative correlation has been measured in social networks \cite{newman2002assortative,hu2009disassortative}.


%

As a third characteristics of complex networks, we evaluate the \textit{local clustering coefficient}, introduced in \cite{watts1998collective}, defined as the ratio of the number of connected pairs of one-hop neighbors to the number of all possible pairs. The clustering coefficient reflects how closed the groups of individuals are. Low-clustering-coefficient nodes are considered to be influential, since their neighbors are isolated from each other \cite{ChenPLOS2013}.
High clustering coefficient is characteristic for WS small-world networks and for random lattice networks. At the same time, the clustering coefficient of Erdos-Renyi random graphs is small. In scale-free networks the clustering coefficient depends on the attachment strategy \cite{barrat2005rate,szabo2003structural}.

\section{Network Model}\label{sec:systemModel}

%
%
%
%

In this paper, we build up a network model of social ties, that considers both
the popularity of the individuals, and their geographical
distribution. Our starting point on the formation of social ties is as follows.

\begin{itemize}
\item Individuals are attracted to connect to popular
  peers\cite{caldarelli2002scale,boguna2003class,perra2012activity,muchnik2013,laniado2011co}.
\item Individuals are less likely to develop social ties with just each other, if there are many others they could connect to \cite{liben2005geographic,scellato2011socio,deville2016scaling}, that is, the emergence of ties depends on the density of the population.
\item The farther away two individuals are from each other, the lower is the probability that they develop social ties \cite{liben2005geographic,lambiotte2008geographical,deville2016scaling,backstrom2010find,goldenberg2009distance,scellato2011socio}.
\item In some networking scenarios, like for infrastructure-less and opportunistic networks, the direct connection between individuals is limited by the used radio technology.
\end{itemize}
Accordingly, the networking scenario is characterized by the
popularity of the nodes, and their geographic location. Following
\cite{laniado2011co,muchnik2013,perra2012activity}, we introduce the \textit{popularity factor} $m$,
and consider that $m$ follows a power-law distribution, that is, $\rho (m)$, the probability that an individual has popularity $m$ is
\begin{equation}
  \rho (m) \sim m^{-\beta },
\end{equation}
where $\beta>2$ and $m \geq m_0$.

As it is required that $\int_{m_0}^{\infty } c m^{-\beta } \, dm=1$, $\rho (m)$ becomes
\begin{equation}\label{powerlawpdf}
  \rho (m) = \frac{\beta -1}{m_0^{1-\beta }}*m^{-\beta }.
\end{equation}

The condition $\beta>2$ ensures the existence of a finite mean popularity value, it also complies with the typical values measured in various networked structures \cite{barabasi2016network}.

We assume that individuals are spread out on the two dimensional space according to a homogeneous Poisson point process (PPP), as it is motivated in \cite{lu2015}. That is, each individual is located independently and uniformly in the two dimensional space, with a density $\lambda$. In any area ${A}$, the number of individuals follows a Poisson distribution with parameter $A\lambda$.
Then, to represent even the popularity, we define the individuals as a homogeneous marked Poisson point process \cite{baccelli2010stochastic}
\begin{equation}\label{mppp}
  {\Phi}=\sum_{i} {\varepsilon_{(x_i,m_i)}},
\end{equation}
with density $\lambda$, where $\varepsilon$ is the Dirac measure on
the Cartesian product $\mathbb{R}^{2} \times \mathbb{R}^{+} $. In point
process \eqref{mppp} $x_i \in \mathbb{R}^2$ denotes the
geographic location of individual $i$ and the associated mark $m_i$ with
probability density function defined in \eqref{powerlawpdf} denotes
the popularity factor of individual $x_i$.

Though here $\Phi$ is
defined as point measure, according to conventions,
we also use $\Phi$ to denote the collection of individuals with
popularity factors. For instance, we use $(x, m) \in \Phi$ ($x\in
\mathbb{R}^2, m \in \mathbb{R}^{+}$) to
denote that the individual located at position $x$ with popularity $m$
belongs to the process $\Phi$. Alternatively, when interpreting $\Phi$
we can say $\Phi((x,m))=1$, since $(x,m)$
belongs to $\Phi$. Additionally, $x\in \Phi$ denotes that the individual
located at $x$ belongs to $\Phi$ whatever the individual's popularity
factor is. For $A \subset \mathbb{R}^2$, $\Phi(A) = \Phi(A\times
\mathbb{R}^{+})$ is used to denote the windowed process where only the
individuals located within the area $A$ are considered.

As a next step, we define the rules that govern the emergence of
social ties, considering i) the willingness to form a tie, and ii) the feasibility of forming a social tie.

First, we build up a gravity law \cite{krings2009urban,scellato2011socio,barthelemy2011spatial} based attachment rule as follows.

To express the effect of the population density, we
define the rank between individuals $i$ and $j$, as the
mean number of individuals that are closer to individual $i$ than
individual $j$ is,

\begin{equation}
  \mathscr{R}_{x_i, x_j} =E[\left| \left\{w: \norm{x_i-w} <
      \norm{x_j-x_i}, w, x_i, x_j \in \Phi \right\}\right|],
\end{equation}
where $\norm{\cdot}$ is Euclidean norm and $\left| \cdot \right|$ returns the number of elements of a given set. Since individual positions follow a homogeneous PPP, the rank of any given pair of individuals is symmetric, and we have
\begin{align}\label{eq:rank}
  \mathscr{R}_{x_i,x_j}&=\mathscr{R}_{x_j,x_i} \nonumber \\ &=E\left[\int_{\norm{x-x_i}<\norm{x_j-x_i}}\!\!\!\!\!\!\!\!\!\!\!\!\!\!\!\!\!\!\!\!\!\!\!\!\!
                                                              \Phi
                                                              (dx) ~~~~~\right] \nonumber \\
                       &=\pi \lambda \norm{x_i-x_j}^2, \forall x_i, x_j \in {\Phi}.
\end{align}
See the details in Appendix A of the supplemental material.


Now we can introduce the \textit{mutual interest factor} $\mathscr{M}_{x_i,x_j}$, the analogue of force in Newton's gravity law, that combines the influence of node popularity and rank.
%
\begin{equation}\label{mutualInterest}
  \mathscr{M}_{x_i,x_j}= \frac{m_i m_j}{\mathscr{R	}^{\alpha }_{x_i,x_j}},
\end{equation}
where $m_i$ and $m_j$ are the popularity factor of individuals
$x_i$ and $x_j$, respectively and parameter $\alpha \geq 0$ is the
\textit{rank exponent}, that reflects the level of influence of the
geographic distance and population density.


Then, based on the mutual interest factor, we introduce the criterion of neighbor relationship, a deterministic rule of developing social ties, using the threshold $\theta$, and consider that two individuals forms a social tie if the mutual interest factor is higher than threshold $\theta$
\begin{equation}\label{criterion1}
  \mathscr{M}_{x_i,x_y} > \theta.
\end{equation}

Independently from the willingness to form social ties, we consider
also the physical limitations of direct connections in a wireless
network. We follow the disk model \cite{gilbert1961random } to define the \textit{accessibility radius} $\delta$, and allow the formation of a social tie between individuals $i$ and $j$ only if the geographic
distance between $x_i$ and $x_j$ is lower than $\delta$
\begin{equation}\label{criterion2}
  \norm{x_i-x_j} < \delta.
\end{equation}

Note that the constrain of accessibility radius can be easily removed by setting $\delta
\rightarrow \infty$, while the effect of the node distance on the mutual interest factor can be eliminated by setting $\alpha = 0$. This allows us to consider the two kinds of spatial effects separately.

\section{Characterization of Geo-social \\ Networks}\label{modelAnalyses}

In this section we derive the main characteristics of geo-social networks that form according to the rules defined in Section \ref{sec:systemModel}. We derive approximate results on the degree distribution, characterize the neighbors of an individual, and derive the local clustering coefficient. We validate the results via simulations in Section \ref{sec:num_results}.

According to Palm theory \cite{baccelli2010stochastic}, under the homogeneous PPP assumption, the statistical characterization of the network can be obtained by analyzing one point of $\Phi$, which we call the typical individual $(x,m)$.



\subsection{Degree Distribution}\label{sub_degreeDistribution}

We define the individuals sharing one-hop social tie with the typical individual $(x,m)$ as neighbors of $(x,m)$.
The neighbor relationship is established by the criterion in
~\eqref{criterion1}, i.e. an individual $(x_n, m_n) \in \Phi$ is able to
set up social tie and be neighbor with $(x, m)$ if their mutual
interest factor $\mathcal{M}_{x, x_n}$ is above threshold $\theta$.

The neighbors of $(x, m)$ form a new point process $\mathbb{N}_{x,m}$, that is generated from $\Phi$, by preserving a point $x_i \in  \Phi$ only if condition
$\mathscr{M}_{x,x_i}>\theta|(x,m)$ is met. That is, $\mathbb{N}_{x,m}$ can be defined as
\begin{equation}\label{neighbourPointProcess}
  \mathbb{N}_{x,m}=\sum_{(x_i,m_i) \in \Phi} \varepsilon_{(x_i,m_i)} \cdot \mathds{1}(\mathscr{M}_{x,x_i}>\theta|(x,m)),
\end{equation}
where the neighbor criterion is explicitly shown by the indicator
function $\mathds{1}(\mathscr{M}_{x,x_i}>\theta|(x,m))$, with
indicator function $\mathds{1}(\cdot)$, $\mathds{1}(True)=1$ and
$\mathds{1}(False)=0$.

To consider finite $\delta$, denote the disk area centred at $x$ with
radius $\delta$ as $B_x(\delta)$,
\begin{equation}
  B_x(\delta)= \{w: \norm{x-w} < \delta \}.
\end{equation}
Then $\Phi(B_x(\delta))$ is the windowed point process $\Phi$
over the area $B_x(\delta)$, and
$\mathbb{N}_{x,m}(B_x(\delta))$ is the random variable giving the number of neighbors of the typical node $(x,m)$
in $B_x(\delta)$.


\begin{lemma}\label{th:lemma1}
  In a random network characterized by marked Poisson point process ${\Phi}=\sum_{i} {\varepsilon_{(x_i,m_i)}}$
  with intensity $\lambda$, the generating function of number of
  neighbors of individual $(x,m)$, within $B_x(\delta)$, with neighbor criterion
  $\mathscr{M}_{x,x_n} > \theta$, where $(x_n, m_n) \in \Phi(B_x(\delta))$, is
  \begin{align}\label{eq:lemma1}
    &G_{\mathbb{N}_{x,m}(B_x(\delta))}(z) \nonumber \\=& e^{\pi  \delta ^2 \lambda
                                                         (Prob[\mathscr{M}_{x,x_n}>\theta|(x,m)] z
                                                         -Prob[\mathscr{M}_{x,x_n} >\theta|(x,m)])}.
  \end{align}
  
\end{lemma}

\begin{IEEEproof}
  Consider the trial as the typical individual $(x,m)$ tries to build social tie with nodes in $B_x(\delta)$
  by a sequence of binomial experiment $b_n, n=1, 2, \cdots, \Phi(B_x(\delta))$. Thus, we have

  \begin{equation}
    \mathbb{N}_{x,m}(B_x(\delta))= \sum_{n=1}^{\Phi(B_x(\delta))} b_n,
  \end{equation}
  where $b_n = \mathds{1}(\mathscr{M}_{x,x_n}>\theta|(x,m))$. The generating function of $b_n$ is
  \begin{align}\label{geBinomial}
    G_b(z)=& \sum_{j=0}^{\infty} Prob[b_n=j] z^{j} \nonumber \\
    =&  Prob[\mathscr{M}_{x,x_n}>\theta|(x,m)] z +Prob[\mathscr{M}_{x,x_n}\le\theta|(x,m)].
  \end{align}
  Since $\Phi$ is a Poisson point process, the generating function of $\Phi(B_x(\delta))$ is
  \begin{align}\label{gePoisson}
    G_{\Phi(B_x(\delta))} (z)
    = & \sum_{j=0}^{\infty} Prob[\Phi(B_x(\delta))=j] z^j \nonumber \\
    =& \sum_{j=0}^{\infty} \frac{e^{-\pi  \delta ^2 \lambda }
       \left(\pi  \delta ^2 \lambda \right)^j}{j!} z^j \nonumber \\
    =& e^{\pi  \delta ^2 \lambda  (z-1)}.
  \end{align}

  Since $\mathbb{N}_{x,m}(B_x(\delta))$ is a compound distribution of $b_n$ and  $\Phi(B_x(\delta))$, we can express its generating function through the generating functions of $b_n$ and  $\Phi(B_x(\delta))$ given in \eqref{geBinomial} and \eqref{gePoisson} \cite{Virtamo3143}
  \begin{align}\label{degreeGenerating}
    &G_{\mathbb{N}_{x,m}(B_x(\delta))}(z) \nonumber \\
    = &E\left[ z^{\mathbb{N}_{x,m}(B_x(\delta))} \right] \nonumber \\
    = & E\left[ E\left[ z^{\sum_{n=1}^{\Phi(B_x(\delta))} b_n}
        \right]|\Phi\left( B_x(\delta) \right) \right] \nonumber \\
    = & E\left[ G_b(z)^{\Phi\left( B_x(\delta) \right)} \right]
        \nonumber \\
    = & G_{\Phi(B_x(\delta))}(G_b(z))
        \nonumber \\
    =& e^{\pi  \delta ^2 \lambda  (Prob[\mathscr{M}_{x,x_n}>\theta|(x,m)] z -Prob[\mathscr{M}_{x,x_n} > \theta|(x,m)])}.
  \end{align}
  
\end{IEEEproof}

We use then the generating function in Lemma \ref{th:lemma1} to derive $k(m)$, the expected degree of a node with popularity $m$, and $P(k)$, the unconditioned degree distribution in the geo-social network.

For a typical individual $(x,m)$, the expected degree $k$ is given by the first moment of $G_{\mathbb{N}_{x,m}(B_x(\delta))}(z)$ as
\begin{align}\label{meanDegree}
  k(m)=& \frac{d}{dz} G_{\mathbb{N}_{x,m}(B_x(\delta))}(z)
         |_{z=1} \nonumber \\
  =& \pi  \delta ^2 \lambda  Prob[\mathscr{M}_{x,x_n}>\theta|(x,m)].
\end{align}

We follow the first-moment based approximation in \cite{caldarelli2002scale} to derive an approximate unconditioned degree distribution
\begin{align}\label{degreeDistribution}
  P(k) \approx & \rho(k^{-1} {(\pi  \delta ^2 \lambda
                 Prob[\mathscr{M}_{x,x_n}>\theta|(x,m)])}) \nonumber \\ & \cdot \frac{d}{dk} k^{-1}{(\pi  \delta ^2 \lambda  Prob[\mathscr{M}_{x,x_n}>\theta|(x,m)])},
\end{align}
where $k^{-1}$ is the inverse function of
\eqref{meanDegree}. The approximation only applies, when the first moment of the network degree distribution converges, but this is typically fulfilled. Also, according to \cite{caldarelli2002scale}, the approximation is tighter for high number of nodes.
The details of calculating $P(k)$ are presented in Appendix B of the supplemental material.

%

\begin{theorem}
  For a network defined in Section \ref{sec:systemModel}, under $\alpha >0$ and $\alpha\beta > \alpha+1$, \\
  a) the network has a scale free degree distribution with
  \begin{equation}
    \tilde{P}[k]\sim k^{\alpha -\alpha \beta  -1},
  \end{equation}
  when the accessible radius $\delta = \infty$, and \\
  b) it has Poisson degree distribution with
  \begin{equation}
    P(k) = \frac{e^{-\pi \lambda \delta^2 } (\pi \lambda
      \delta^2)^k}{k!},
  \end{equation}
  when $\delta$ is strongly limited, such that $\theta(\pi \lambda \delta^2)^{\alpha} <  m_0^2$.
\end{theorem}

\begin{IEEEproof}
  Let us first derive $Prob[\mathscr{M}_{x,x_n}>\theta|(x,m)]$, needed in
  \eqref{meanDegree} and \eqref{degreeDistribution}. We consider two
  significantly different regimes of network formation. Remember that
  node popularity has a minimum value, $m \geq m_0$. As a consequence,
  for limited accessibility radius $\delta$, $\mathscr{M}_{x,x_n}$, the
  mutual interest factor has a minimum value of $m_0^2/(\pi \lambda
  \delta^2)^{\alpha}$. If the threshold $\theta$ is smaller than this
  value, nodes form social tie to all other nodes within the
  accessibility radius. Otherwise, the existence of the social tie
  between two individuals depends on their popularity and distance, and on the
  population density.

  We first consider the case when $\theta(\pi \lambda \delta^2)^{\alpha} >  m_0^2$.
  In this case the existence of a social tie depends on $m$ and $m_n$.

  We denote by $f_{\delta} (r)$ the distribution function of the distance $r$ between the typical point and a randomly chosen point $x_n \in \Phi(B_x(\delta))$. We know that \cite{baccelli2010stochastic,haenggi2009interference}

  \begin{equation}\label{radialDensity}
    f_{\delta} (r)= \frac{2 r}{\delta^2}, 0 < r < \delta,
  \end{equation}
  where $r=\norm{x-x_n}$ , $(x_n, m_n) \in \Phi(B_x(\delta))$.


  Using \eqref{powerlawpdf} and \eqref{radialDensity}, and under the condition $\alpha\beta > \alpha + 1$, we can write
  \begin{align}\label{connProb}
    &Prob[\mathscr{M}_{x,x_n}>\theta|(x,m)]\nonumber \\
    =&\iint_{\mathscr{M}_{x,x_n}>\theta}\rho(m_n) f_{\delta}(r) dm_n dr \nonumber\\
    =& \frac{\alpha  (\beta -1) }{\pi  \delta ^2 \lambda  (\alpha  \beta -\alpha-1)} \left(\frac{m_0 m}{\theta  }\right)^{1/\alpha }- o(\delta,m),
  \end{align}
  where
  \begin{align}\label{connProbResidual}
    o(\delta,m)
    =& \frac{\pi ^{\alpha -\alpha  \beta } \delta ^{-2 \alpha  (\beta -1)} \lambda ^{\alpha -\alpha  \beta }}{\alpha  \beta -\alpha -1} \left(\frac{m_0 m}{\theta }\right){}^{\beta -1}.
  \end{align}

  The condition ensures that the denominators are positive, and is not too restrictive, considering that $\beta > 2$ is found measurement studies. The details of the derivation of \eqref{connProb} can be found in Appendix C of the supplemental material.

  We can now return to \eqref{meanDegree} and \eqref{degreeDistribution} and express $k(m)$:

  \begin{equation}\label{degreeOnm}
    k(m)= \frac{\alpha  (\beta -1) }{\alpha  \beta -\alpha-1} \left(\frac{m_0 m}{\theta  }\right)^{1/\alpha }- \pi  \delta ^2 \lambda o(\delta,m).
  \end{equation}

  For unlimited tie length, \textit{i.e.} $\delta \rightarrow
  \infty$, \eqref{degreeOnm} can be written as
  \begin{equation}\label{limitDegreeOnm}
    \tilde{k}(m) = \lim\limits_{\delta \rightarrow \infty} k(m)=\frac{\alpha  (\beta -1) }{\alpha  \beta -\alpha-1} \left(\frac{m_0 m}{\theta  }\right)^{1/\alpha }.
  \end{equation}

  Under the same condition of $\delta \rightarrow \infty$, we can express the limiting degree distribution


  \begin{equation}\label{degreeDistributionFianl}
    \tilde{P}(k) \approx \frac{(\alpha  \beta -\alpha )^{\alpha
        \beta -\alpha +1}  \theta ^{1-\beta } m_0^{2 \beta -2}
      k^{\alpha -\alpha \beta  -1}}{(\alpha  \beta -\alpha -1)^{
        \alpha  \beta -\alpha}}, \alpha \beta > \alpha +1.
  \end{equation}

  That is, for given $\alpha$, $\beta$, $\theta$ and $m_0$ values,
  %
  %
  \begin{equation}
    \tilde{P}(k)\sim k^{\alpha -\alpha \beta  -1}, \alpha \beta >
    \alpha +1, \nonumber
  \end{equation}

  Let us now consider the case, when $\theta (\pi \lambda \delta^2)^{\alpha} \leq m_0^2$. Since all $m_n > m_0$, all the nodes locating inside $B_x(\delta)$ connect with the typical node, no matter what the popularity factor of the typical node
  is. That is, we have
  \begin{equation}\label{connProb1}
    Prob[\mathscr{M}_{x,x_i}>\theta|(x,m)]=1, 
  \end{equation}
  and, from \eqref{meanDegree}, the mean number of neighbors becomes
  \begin{equation}\label{meanPoisson}
    k(m) = \pi  \delta ^2 \lambda.
  \end{equation}

  In this case we can derive the degree distribution directly from the generating function $G_{\mathbb{N}_{x,m}(B_x(\delta))}$, which now becomes independent of $m$, and we get


  \begin{align}\label{degreePoisson}
    P(k)&=\frac{1}{k!} \frac{\partial^k
          G_{\mathbb{N}_{x,m}(B_x(\delta))}}{\partial z^k}|_{z=0} \nonumber\\
        &= \frac{e^{-\pi \lambda \delta^2 } (\pi \lambda
          \delta^2)^k}{k!}. 
  \end{align}

\end{IEEEproof}

Finally, let us investigate, what kind of network is formed under the proposed mutual interest factor based attachment, if the effect of node distance is removed, that is, $\alpha = 0$.
\begin{theorem}
  For a network defined in Section \ref{sec:systemModel} and under $\alpha = 0$, the nodes have Poisson degree distribution if $\theta \leq m_0^2$. Otherwise, the node degrees are infinite if $\delta = \infty$, but have power low distribution with a scaling exponent of two if $\delta < \infty$.
\end{theorem}

\begin{IEEEproof}
  In the $\alpha = 0$  case the mutual interest factor is reduced to
  $\mathscr{M}_{x_i,x_j}^0=m_i m_j$, which also means that the accessibility radius $\delta$ does not affect the probability that nodes within accessible distance form social tie or not.

  When $\alpha = 0$ and $\theta \leq m_0^2$, the node degree distribution is Poisson, following the results of  \eqref{meanPoisson} and \eqref{degreePoisson}.

  When $\theta  > m_0^2$, the probability $Prob[\mathscr{M}_{x,x_n}>\theta|(x,m)]$ of the typical node's
  connection to an random neighbor $(x_n,m_n)$ is reduced to
  \begin{align}\label{zeroConnProb}
    &Prob[\mathscr{M}_{x,x_n}^0>\theta|(x,m)]\nonumber \\
    =&\int_{\mathscr{M}_{x,x_n}^0>\theta}\rho(m_n) dm_n \nonumber\\
    \approx&(m_0 m/\theta)^{\beta-1}.
  \end{align}
  The details of the derivation can be found in Appendix D of the supplemental material.

  According to~\eqref{meanDegree}, $k^0(m)$, the average number
  of connected neighbors of typical node $(x,m)$ becomes for the
  zero-rank-exponent case
  \begin{equation}\label{k0m}
    k^0(m) \approx \pi \lambda \delta^2 (m_0 m/\theta)^{\beta-1}.
  \end{equation}

  Note, that without accessibility constraint the average number of connected neighbors of typical node $(x,m)$ is infinite  \textit{i.e.} $\lim\limits_{ \delta \rightarrow \infty}k^0(m)=\infty$. 


  For finite accessible radius $\delta$, $P^0(k)$ can be directly calculated from the closed from equation of $k^0(m)$ in \eqref{k0m}, and \eqref{degreeDistribution}, resulting
  \begin{equation}\label{zeroDegreeDist}
    P^0(k) \approx \frac{\pi \delta^2 \lambda \theta^{1-\beta}m_0^{-2(1-\beta)}} {k^{2}},
  \end{equation}
  that is, if the probability of establishing social ties does not
  decrease with the distance and density, the degree distribution in the network follows a power law, with a scaling exponent of two.
  
\end{IEEEproof}


\begin{remark}
  We can now summarize the effect of the spatial distribution on the
  node degrees. Under large accessibility radius $\delta$, the
  scale-free property of the network is preserved, though the scaling
  exponent is affected by the geographic constraints through
  $\alpha$. The larger the value of $\alpha$ is, the fewer are the
  nodes with very large degree. Moreover, as shown in \eqref{degreeDistributionFianl}, the degree distribution in the geo-social network is independent of $\lambda$, the density of the nodes, reflecting the findings \cite{scellato2011socio}.
  The network characteristics undergo a significant transformation as $\delta$ is decreased, the network loses its scale-free property, and becomes a random lattice network with Poisson degree distribution.
\end{remark}

\subsection{Neighbor Degree}\label{sub:neighborDegree}

Next, we express how the average neighbor degree of an individual depends on its popularity. We consider also the special case of $\alpha = 0$, when we also derive how the neighbor degree depends on the node degree itself, that is, the assortativity of the emerging networks. The network assortativity for the general case of $\alpha > 0$ will be evaluated via simulations.

Let us denote by $\mathbb{N}_{x,m}^s (B_x(\delta))$ the sum of the popularity of the neighbors of the typical individual, within the accessibility radius $\delta$, by $\bar{m}_n$ the the average popularity factor and by $\bar{k}_{nn}(m)$ the average degree of the neighbours of the typical individual $(x,m)$. Finally we denote by $\bar{k}_{nn}(k)$ the average degree of neighbors of nodes with degree $k$.

\begin{theorem}\label{th:neighbor-deg1}
  For a network defined in Section \ref{sec:systemModel}, and under $\alpha >0$ and $\alpha\beta > \alpha+1$,
  the average neighbor degree of a node with popularity $m$ can be approximated as
  \begin{equation}\label{avNeighborDeg}
    \bar{k}_{nn}(m)\approx\frac{\alpha  (\beta -1) }{\alpha
      \beta -\alpha-1} \left(\frac{m_0 \bar{m}_n}{\theta
      }\right)^{1/\alpha }- \pi  \delta ^2 \lambda
    o(\delta,\bar{m}_n), \alpha>0,
  \end{equation}
  where $\bar{m}_n$ is
  \begin{align}\label{meanPF}
    \bar{m}_n & \approx \\
              & \left \{
                \begin{array}{rcl}
                  &-\frac{(\beta -1) \pi ^{-\alpha  (\beta -2)} m_0^{\beta -1} \delta
                    ^{-2 \alpha  (\beta -2)} \lambda ^{-\alpha  (\beta -2)}
                    \left(\frac{m}{\theta }\right)^{\beta -2}}{(\beta -2) (\alpha
                    (\beta -2)-1)Prob[\mathscr{M}_{x,x_n}>\theta|(x,m)]} \\
                  &+ \frac{\alpha  (\beta -1) m_0^{\frac{1}{\alpha }+1} \left(\frac{m}{\theta }\right)^{1/\alpha }}{\pi  \delta ^2 \lambda  (\alpha  (\beta -2)-1)Prob[\mathscr{M}_{x,x_n}>\theta|(x,m)]}    , {m<\frac{\theta (\pi \lambda \delta^2)^{\alpha}}{m_0}},
                  \\
                  &\frac{ m_0(\beta-1)}{\beta-2}, {m>\frac{\theta (\pi \lambda \delta^2)^{\alpha}}{m_0}.}
                \end{array} \right. \nonumber
  \end{align}
  The second case with condition $m>\frac{\theta (\pi \lambda \delta^2)^{\alpha}}{m_0}$ is the case when the typical individual is connected to all others in its accessibility region, and as expected, the neighbors' average popularity is independent from $m$.
\end{theorem}

\begin{IEEEproof}
  We approximate the average neighbor popularity factor $\bar{m}_n$ as

  \begin{align}\label{meanPFstep1}
    \bar{m}_n & = E \left [ \frac{\mathbb{N}_{x,m}^s (B_x(\delta))}{\mathbb{N}_{x,m} (B_x(\delta))}\right ] \approx\frac{E[\mathbb{N}_{x,m}^s (B_x(\delta))]}{k(m)},
  \end{align}
  where $k(m)$ is given by \eqref{meanDegree}.

  The sum of the popularity of the neighbors can be expressed as

  \begin{equation}\label{neighbourSumPop}
    \mathbb{N}_{x,m}^{s} (B_x(\delta))=\sum_{(x_n,m_n) \in \Phi}
    \varepsilon_{(x_n,m_n)} \cdot
    \mathds{1}(\mathscr{M}_{x,x_n}>\theta|(x,m)) \cdot m_n,
  \end{equation}
  and its mean value can be approximated as \cite{servedio2004vertex}
  \begin{equation}\label{meanSumNeighborDeg}
    E[\mathbb{N}_{x,m}^s (B_x(\delta))] \approx \pi\delta^2\lambda \iint_{\mathscr{M}_{x,x_n}>\theta} m_n \rho(m_n)f_{\delta}(r) dm_n dr.
  \end{equation}

  $E[\mathbb{N}_{x,m}^s (B_x(\delta))]$ can be expressed in closed form and by replacing it in \eqref{meanPFstep1}, \eqref{meanPF} follows.


  Finally, we arrive to \eqref{avNeighborDeg} by substituting ~\eqref{meanPF} into \eqref{degreeOnm}.


\end{IEEEproof}

This result is hard to interpret, and therefore in Section \ref{sec:num_results} we evaluate
the results numerically, also assessing the effect of the approximations.

For comparison, we present the final results on $\bar{k}_{nn}^0(m)$, the average
neighbor degree for $\alpha=0$.

\begin{theorem}\label{th:neigbor-degree-a0}

  For the network defined in Section \ref{sec:systemModel}, but under $\alpha=0$, that is, when the geographic distance does not affect the mutual interest factor, the average neighbor degree of a node with popularity $m$, $\bar{k}_{nn}^0(m)$ is approximated by constant $\pi \lambda \delta^2$ when $\theta<m_0^2$. Otherwise, when $\theta > m_0^2$, we have
  \begin{align}\label{k0nnm}
    &\bar{k}_{nn}^0(m)\approx\left\{
      \begin{array}{rcl}
        & \pi  \left(\frac{\beta -2}{\beta -1}\right)^{1-\beta } \delta ^2
          \lambda  \theta ^{1-\beta } m_0^{2 \beta -2} , {m> \theta/m_0}\\
        &\pi  \left(\frac{\beta -2}{\beta -1}\right)^{1-\beta } \delta ^2
          \lambda  m_0^{\beta -1} m^{1-\beta }, {m<\theta/m_0},
      \end{array} \right.
  \end{align}
  and the average neighbor degree as a function of the degree $k$ can be expressed as
  \begin{align}\label{meanNeigbor0}
    &\bar{k}_{nn}^0(k)\approx   &\frac{\left( \pi \delta ^2 \lambda \right)^2}{k} \left(\frac{\beta -2}{\beta -1}\right)^{1-\beta } \left( \frac{m_0^2}{\theta} \right)^{\beta-1}, {k \leq \pi \delta ^2 \lambda}.
  \end{align}
  Note, that the constraint $k \leq \pi \delta ^2 \lambda$ comes from the limit on the average number of nodes available in an area.
  
\end{theorem}

\begin{IEEEproof}
  When $\theta<m_0^2$, a node can connect with any other nodes
  within its accessability radius and thus $\bar{k}^{0}_{nn}$ can be
  approximated by $\pi \lambda \delta^2$. For $\theta>m_0^2$ case, we have closed form expression of the mean degree and
  the degree distribution of the typical node $(x,m)$ in~\eqref{k0m}
  and \eqref{zeroDegreeDist}. \eqref{k0m} gives the expected number of connected neighbors. Applying
  $\alpha=0$ in \eqref{meanPFstep1} gives the $\bar{m}_n$ for
  this case and substituting the result in \eqref{k0m} gives average
  node degree of a node with popularity $m$ as shown in \eqref{k0nnm}. Accordingly, \eqref{meanNeigbor0}, where
  $\bar{k}_{nn}^{0}$ is as function of $k$ is obtained by combining
  \eqref{k0m} and \eqref{k0nnm} to eliminate $m$.
  The details of the derivations can be found in Appendix E of the supplemental material.


\end{IEEEproof}

\begin{remark}
  We can now summarize the analytic findings on neighbor degrees in the geo-social network. Theorem \ref{th:neighbor-deg1} gives the neighbor degree as a function of the node popularity. The expression is too complex to give direct conclusions, and therefore will be analysed numerically, together with the node degree -- neighbor degree relationship.

  Theorem \ref{th:neigbor-degree-a0} discusses the $\alpha = 0$ case and shows that the neighbor degree is uncorrelated with the node popularity, unless $m<\theta/m_0$, that is, for small popularity values. Under $\alpha = 0$ network assortativity can be analysed as well, where the results show that the network is disassortative at small $k$ values, and neutral otherwise.


\end{remark}

\subsection{Clustering Coefficient}\label{sub:ClusterCoeff}

Finally, we evaluate the local clustering coefficients of the geo-social network nodes, defined by the probability that two neighbors of a node are also connected by a social tie.

We denote the event that node $(x_i,m_i)$ connects with $(x_j,m_j)$ by $\mathscr{L}_{(x_i,m_i),(x_j,m_j)}$, where $i \neq j$. Then, the clustering coefficient ${C}(x,m)$ of
the typical node $(x,m)$ in our social network model can be expressed as the conditional probability of
$\mathscr{L}_{(x_i,m_i),(x_j,m_j)}$, given that $(x_i,m_i)$ and $(x_j,m_j)$ are neighbors of $(x,m)$, that is, $\mathscr{L}_{(x,m),(x_i,m_j)}$ and $\mathscr{L}_{(x,m),(x_j,m_j)}$ holds,
\begin{align}\label{clusterCoefficient}
  &{C}(x,m) \nonumber \\
  &= Prob[\mathscr{L}_{(x_i,m_i),(x_j,m_j)}| \mathscr{L}_{(x,m),(x_i,m_i)}, \mathscr{L}_{(x,m),(x_j,m_j)}] \nonumber\\
  &=\frac{Prob[\mathscr{L}_{(x_i,m_i),(x_j,m_j)},\mathscr{L}_{(x,m),(x_i,m_i)}, \mathscr{L}_{(x,m),(x_j,m_j)}]}{Prob[\mathscr{L}_{(x,m),(x_i,m_i)}, \mathscr{L}_{(x,m),(x_j,m_j)}]}.
\end{align}

Since the location of the nodes follows a Poisson point process, and the popularity factors are assigned to the nodes independent from their locations, the denominator of \eqref{clusterCoefficient} contains independent events and can be expressed as

\begin{align}\label{clusterDenominator}
  &Prob[\mathscr{L}_{(x,m),(x_i,m_i},\mathscr{L}_{(x,m),(x_j,m_j)}]
    \nonumber \\
  &=\left( Prob[\mathscr{L}_{(x,m),(x_i,m_i}] \right)^2 \nonumber \\
  &=\left( Prob[\mathscr{M}_{x,x_i}>\theta|(x,m)] \right)^2 \nonumber \\
  &=\left\{
    \begin{array}{l}
      1, \quad {\theta(\pi \lambda \delta^2)^{\alpha}      <      m m_o}\\
      \left(  \frac{\alpha  (\beta -1) }{\pi  \delta ^2 \lambda  (\alpha
      \beta -\alpha-1)} \left(\frac{m_0 m}{\theta  }\right)^{1/\alpha }- o(\delta,m) \right)^2, \\
      \quad\quad {\theta(\pi \lambda \delta^2)^{\alpha}     >      m m_o},
    \end{array} \right.
\end{align}
where the second step applies \eqref{connProb} and \eqref{connProb1}.


On the other hand, we need to leave the numerator in integral form, that is
\begin{align}
  & Prob[\mathscr{L}_{(x_i,m_i),(x_j,m_j)},\mathscr{L}_{(x,m),(x_i,m_i)},
    \mathscr{L}_{(x,m),(x_j,m_j)}]
    \nonumber \\
  = & \underset{\mbox{\tiny$\begin{array}{c}
                              \mathscr{M}_{x,x_i}>\theta\\
                              \mathscr{M}_{x,x_j}>\theta\\
                              \mathscr{M}_{x_i,x_j}>\theta
                            \end{array}$}}{\iint\iint\iint}
  \rho(m_i) f_{\delta}(r_i)
  \rho(m_j) f_{\delta}(r_j) dm_i dr_idm_j dr_j d\eta_i d\eta_j,
\end{align}
where $r_i=\norm{x-x_i}$, $r_j=\norm{x-x_j}$, $\eta_i$
and $\eta_j$ are the relative angles of node $(x_i,m_i)$ and
$(x_j,m_j)$ against $(x,m)$ respectively in polar coordinate.

We will evaluate $C(x,m)$ for general $m$ numerically. The following theorem addresses the local clustering coefficient values of high popularity nodes.

\begin{theorem}\label{th:clustering-limit}
  Consider a geo-social network as given in Section \ref{sec:systemModel}, and consider the high popularity nodes with $m > \theta(\pi \lambda \delta^2)^{\alpha}/m_0$. The clustering coefficient of these nodes is independent from the popularity value $m$.
\end{theorem}

\begin{IEEEproof}
  When $m > \theta(\pi \lambda \delta^2)^{\alpha}/m_0$, the typical node $(x,m)$ is connected to all other nodes in its reachability area $B_x(\delta)$. That is, $Prob[\mathscr{L}_{(x,m),(x_i,m_i)}]=1$ for all $x_i$ in $B_x(\delta)$ and \eqref{clusterCoefficient} is reduced to
  \begin{align}\label{constantCluster}
    C(x,m) &=  Prob[\mathscr{L}_{(x_i,m_i),(x_j,m_j)}] \nonumber \\
           &=\iiint_{\mathscr{M}_{x_i,x_j}>\theta} f(l,\delta) \rho(m_i) \rho(m_j) dl dm_i dm_j,
  \end{align}
  where $l=\norm{x_i-x_j}$ denotes the distance between any $(x_i,m_i)$ and $(x_j,m_j)$ in
  $B_x(\delta)$. Under independent node locations the probability density function of $l$ is known \cite{MOLTCHANOV2012}
  \begin{equation}\label{eq:dens-l}
    f(l,\delta)=\frac{2 l}{\delta ^2}\left(\frac{2} {\pi }\cos ^{-1}\left(\frac{l}{2
          \delta }\right)-\frac{l }{\pi  \delta }\sqrt{1-\frac{l^2}{4 \delta
          ^2}}\right).
  \end{equation}

  Since now \eqref{eq:dens-l} and consequently also \eqref{constantCluster} is independent from $m$, the theorem follows.

\end{IEEEproof}

\begin{remark}
  To summarize, for general values of $m$, the local clustering coefficient needs to be calculated numerically. Theorem \ref{th:clustering-limit} however shows that the clustering coefficient of high popularity nodes is independent from their popularity value $m$. From \eqref{constantCluster} we see that this limiting clustering coefficient depends on the network parameters through $\mathscr{M}_{x_i,x_j}$. Therefore, we expect that the limiting clustering coefficient decreases with decreasing $\mathscr{M}_{x_i,x_j}$, that is, with increasing accessability radius $\delta$ and rank exponent $\alpha$. We will evaluate the clustering coefficient as a function of the node degree via simulations.
\end{remark}

\section{Numerical Results}
\label{sec:num_results}

In this section we validate the analytic results derived in Section
\ref{modelAnalyses} via simulations, and discuss the properties of the
emerging geo-social networks. We consider a typical scaling exponent $\beta=2.5$, \cite{barabasi2016network} and $\alpha=1$, which results in mutual interest that decreases with the square of the node distance \cite{lambiotte2008geographical,deville2016scaling}. We consider a node density of $\lambda = 10^{-4}$, and to validate the results with infinite accessibility radius we consider $\delta = 2000$, which allows more than one thousand possible neighbors.
Parameter $m_0$ is the normalization constant for the popularity distribution, and we select $m_0=10$ to allow even low popularity values. We select $\theta$ values so that they limit the social tie formation. When $\alpha > 0$ the default value is $\theta = 10$.
%
Unless specifically shown, parameters of simulations
are taken according to these default values. In each simulation experiment, we consider a typical node and all the nodes within its accessibility region $\delta$. We place the nodes according to a Poisson point process, and select popularity values following \eqref{powerlawpdf}. 


\begin{figure}
  \centering
  \includegraphics[scale = 0.3]{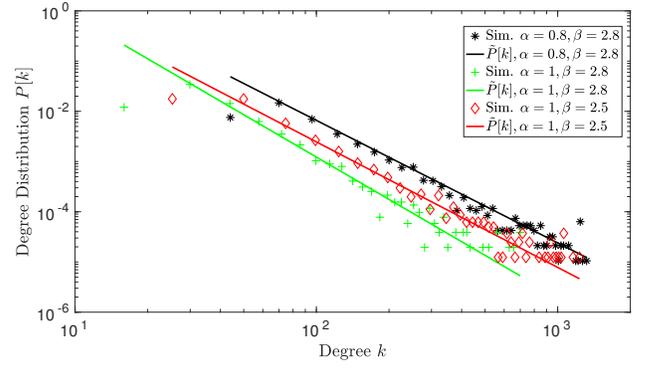}\\
  \caption{Degree distribution in the generated social network for different $\alpha$ and $\beta$ parameters. }\label{fig:degreeDistribution}
\end{figure}

\subsection{Degree Distribution}

We first evaluate the degree distribution of the emerging networks. For each considered parameter combination, we generate 5000 random topologies. 

\begin{figure}
  \centering
  \includegraphics[scale = 0.3]{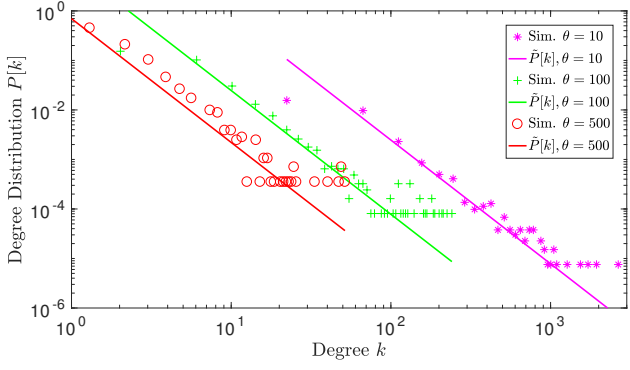}
  \caption{Degree distribution of the generated social network under different \textit{mutual interest factor} threshold $\theta$.} \label{fig:influenceOfThreshold}
\end{figure}

Figs~\ref{fig:degreeDistribution} and  \ref{fig:influenceOfThreshold} show the degree distribution in the $\delta \rightarrow \infty$ case, and evaluate the effect of the scaling exponent $\beta$, the rank exponent $\alpha$, and the mutual interest limit $\theta$.

We show the theoretic results of $\tilde{P}(k)$ in \eqref{degreeDistributionFianl} for infinite accessibility radius, and compare it to simulation results with finite, large $\delta = 2000$.
Fig.~\ref{fig:degreeDistribution} shows that the $\tilde{P}(k)$ approximation provides accurate results, and validates that the degree distribution in the geo-social network follows a power law. The $\alpha$ and $\beta$ parameters have some effect on the exponent, but this effect is rather low in the considered, realistic range of parameters.
The figure shows that larger rank exponent $\alpha$ leads to lower
degrees, since nodes in the same distance need higher popularity to
develop social ties. On the other hand, larger $\beta$ decreases
the probability of the existence of high popularity factor nodes in
the network and the change of generated social-tie topology is consistent with the change of $m$'s distribution.


Fig.~\ref{fig:influenceOfThreshold} evaluates how $\theta$, the mutual
interest factor threshold influences the degree distribution of the
developed networks. The variation of threshold $\theta$ makes no difference
in the exponent of degree distribution of the developed networks,
which is consistent with our analytical conclusion in~\eqref{degreeDistributionFianl}.
By comparing the curves under different values of $\theta$, we see its impact on the
range of the degree of the developed networks. As $\theta$ is increased, fewer node pairs pass the threshold,
the node degrees decrease, the network becomes sparser. In the simulations, the maximum degree is limited due to the necessary limit on $\delta$.


\begin{figure}
  \centering
  \subfigure[Influence of the accessibility radius, when $\alpha = 1$]{\includegraphics[scale =
    0.3]{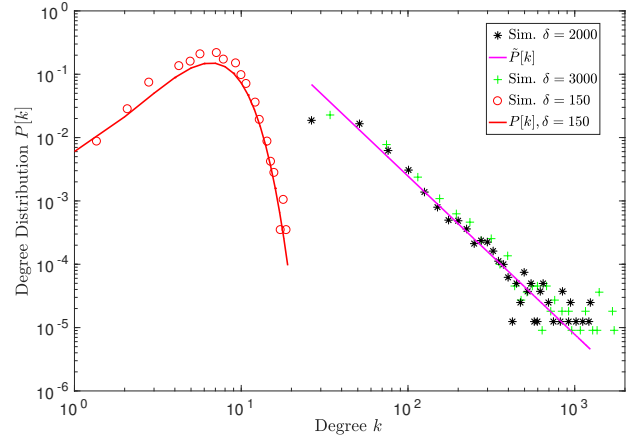}}
  \centering
  \subfigure[Influence of accessibility radius $\delta$ without
  geographic impact, that is, $\alpha = 0$ and  $\theta=10$.]{\includegraphics[scale=0.3]{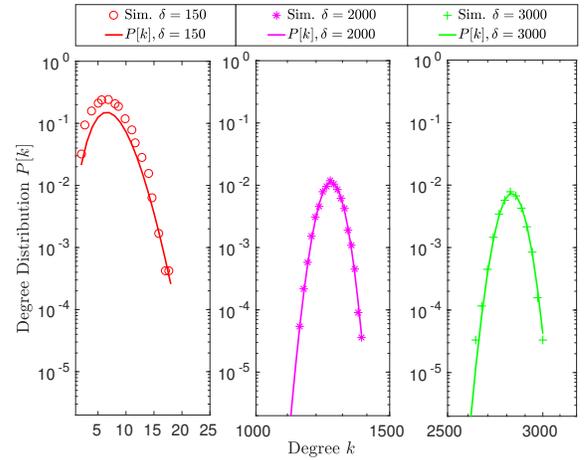}}
  \centering
  \subfigure[Influence of accessibility radius $\delta$ without
  geographic impact, that is, $\alpha = 0$ and restrictive $\theta=1000$.]{\includegraphics[scale=0.3]{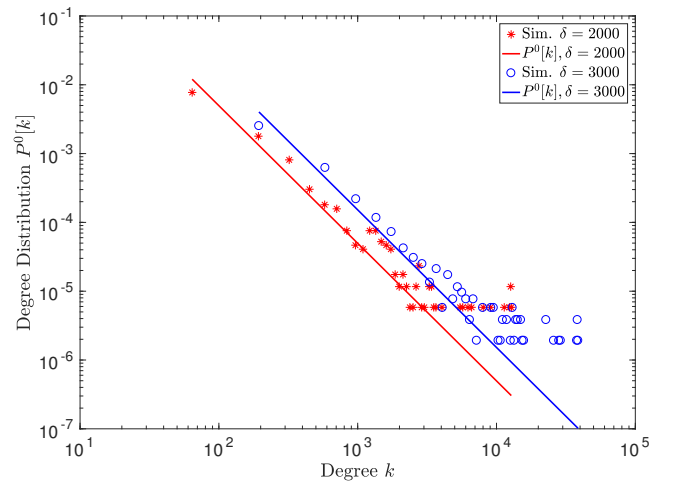}}
  \caption{Degree distribution in the generated social networks under different values of the accessibility radius $\delta$.}\label{fig:influenceOfDelt}
\end{figure}


Fig.~\ref{fig:influenceOfDelt}-(a) shows the degree distributions
of the generated social-tie topology under varying $\delta$ values and $\alpha=1$ comparing simulation results with
analytic results for $\delta \rightarrow \infty$ in \eqref{degreeDistributionFianl} and for limited $\delta$ in \eqref{degreePoisson}.
The figure shows that the degree distribution approaches the $\delta \rightarrow \infty$ limit already under reasonable $\delta$ values. For small accessibility radius, the figure confirms the analytic result that the degree distribution becomes Poisson.

Figs~\ref{fig:influenceOfDelt}-(b),(c) evaluates the node degree distribution when the mutual interest factor is not dependent on the distance, that is, $\alpha = 0$. 

In Section \ref{sub_degreeDistribution} we have seen that the ranges of power law and Poisson degree distribution are determined by the relationship of $\theta$ and $m_0^2$, and are independent of $\delta$. Fig. ~\ref{fig:influenceOfDelt}-(b) shows the degree distribution for the default $m_0$ and $\theta=10$ values. The network remains Poisson, and $\delta$ affects only the average node degree. Fig. ~\ref{fig:influenceOfDelt}-(c) shows the degree distribution when $\theta$ is increased to $1000$, such that $\theta > m_0^2$. The network in this case is scale-free, and the simulation results verify the approximation of \eqref{zeroDegreeDist}.

To summarize the effects of the node distances, we can conclude that
distance dependent attachment, governed by the rank exponent $\alpha$
and the accessibility radius $\delta$ allows the formation of scale
free networks, when
$\delta$ are not too small, otherwise,
the network becomes Poisson, like a random lattice. Under
$\alpha=0$, the mutual interest factor is independent from the node
distances and density, and the network structure is determined by the connection threshold $\theta$.


\subsection{Neighbor Degree}

\begin{figure}
  \centering
  \subfigure[Average degree of neighbors against
  popularity factor and degree of typical node. ($\delta=2000$)
  ]{\includegraphics[scale=0.3]{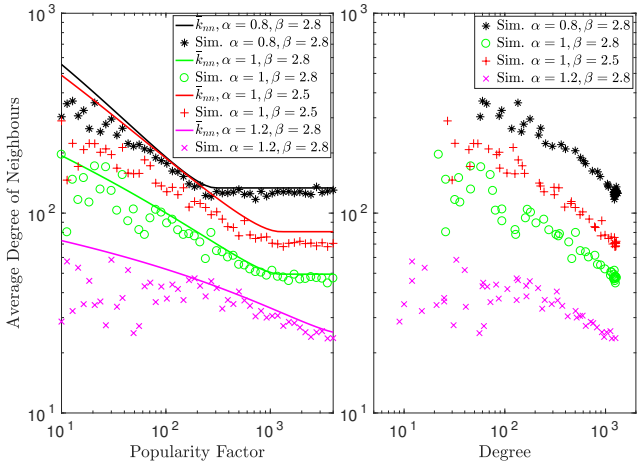}}
  \centering
  \subfigure[Average degree comparison of neighbors between large and
  small accessible radius]{\includegraphics[scale=0.3]{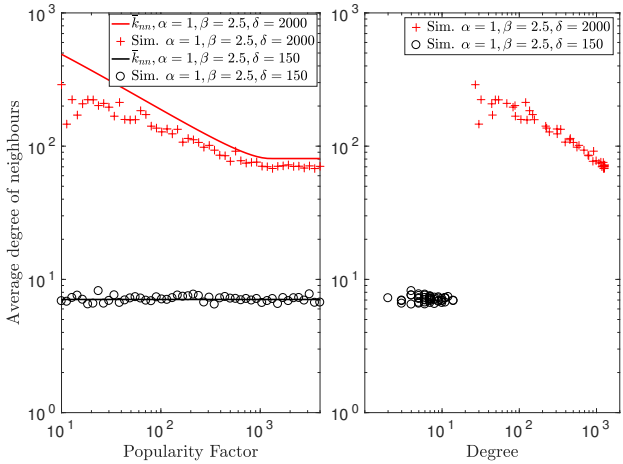}}
  \centering
  \caption{Degree of neighbours as a function of node popularity
    and degree for $\alpha>0$}\label{fig:neighborsDegree}
\end{figure}

Next, we validate the analytic results on the neighbor degree distribution from Section \ref{sub:neighborDegree}, and evaluate the results numerically. For the simulation results, we sample the popularity of the typical node according to the logarithmic scale, and perform 10 simulations with random topology for each considered $m$ value.

First we consider $\alpha > 0$.


Fig.~\ref{fig:neighborsDegree}-(a) shows the average degree of the neighbors as the function of the node popularity and as the function of the node degree. Considering node popularity, we compare the approximate results $\bar{k}_{nn}$ from~\eqref{avNeighborDeg} to simulation results. 

The average neighbor degree decreases with the increase of the popularity factor. The reason for this decrease is that low popularity nodes can rarely connect to each other. As predicted by~\eqref{meanPF},
the average neighbor degree becomes independent from the popularity of the node itself at
high popularity factor values, once the node is able to develop connections with all other nodes within
its accessibility radius. For a given $m$, increased $\alpha$
decreases the average neighbor degree, since the few large degree
nodes may become to be too far away to connect to. Similarly, larger
$\beta$  decreases the density of high popularity nodes, leading as
well to decreased average neighbor degree. Considering assortativity, we see that the negative correlation is preserved, that is, the network is disassortative. This is reasonable, since there is strong positive correlation between popularity and node degree. The network is disassortative for all $k$ values, since for high $m$, $k$ is limited by the accessability radius $\delta$.

Fig.~\ref{fig:neighborsDegree}-(b) shows in addition the case of small accessibility radius, where the network becomes neutral, that is, the popularity of a node does not affect the average popularity of its neighbors, as predicted in~\eqref{avNeighborDeg}. Due to the same reason, the expected average neighbor degree is independent of $k$.

We can conclude, that in general, the emerging geo-social networks are disassortative, with parameters that depend on the geographic limitations.
At very low accessibility radius the network becomes neutral, resembling a random lattice.

\begin{figure}
  \centering
  \includegraphics[scale = 0.3]{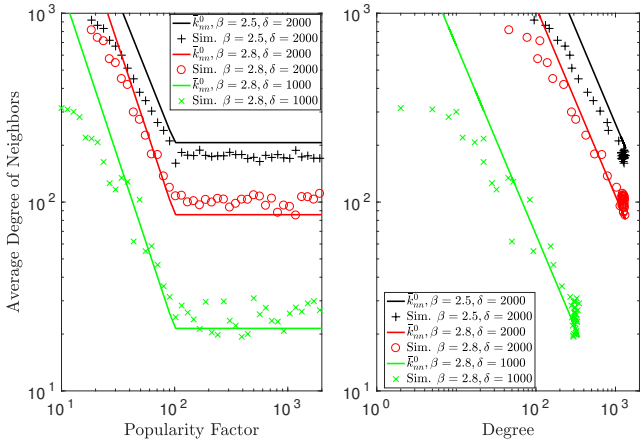}\\
  \caption{Degree of neighbours as a function of node popularity and degree for $\alpha=0$.}\label{fig:knn0}
\end{figure}

Finally, Fig.~\ref{fig:knn0} considers the $\alpha = 0$, that is, when the distance does not affect the formation of social ties. We set $\theta = 1000$, in that case the degree distribution was power law. For this scenario analytic results are available from \eqref{k0nnm} and \eqref{meanNeigbor0}. The neighbor degree decreases with the popularity up to the threshold value in \eqref{k0nnm}. The network is disassortative, as expected from \eqref{meanNeigbor0}. 

Considering the accuracy of the approximations on Figs~\ref{fig:neighborsDegree}-\ref{fig:knn0}, we see that the approximation is accurate at high popularity factor and high node degree, while it overestimates $k_{nn}$ otherwise. The reason is the mean popularity value based degree approximation in \eqref{avNeighborDeg}, which then disregards the degree limitation of very high popularity nodes.

We can conclude that the geographic distance affects the neighbor degree distribution. Considering low popularity nodes, the neighbor degree decreases with increasing popularity, but at high popularity values it settles to a constant, due to the spatial constraints. Since these constraints limit the node degree itself, the networks are disassortative in the entire region of node degrees. At very low accessibility radius the network becomes neutral, resembling a random lattice.

\subsection{Clustering Coefficient}


\begin{figure}
  \centering
  \includegraphics[scale = 0.3]{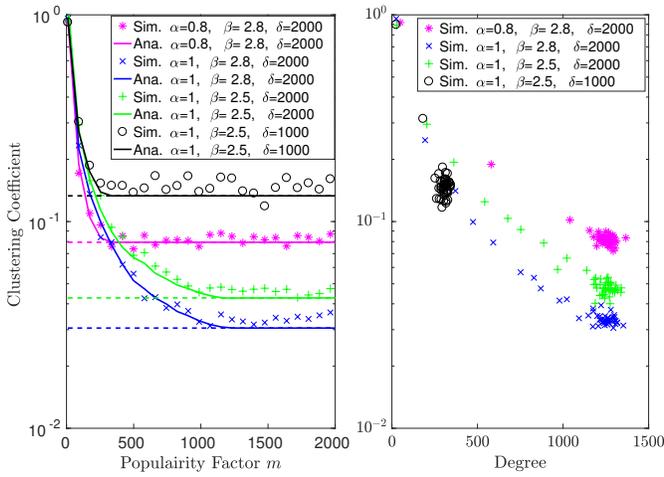}\\
  \caption{Clustering coefficient as a function of the node popularity and degree for $\alpha >0$. }\label{fig:cluster}
\end{figure}

Section \ref{sub:ClusterCoeff} models the local clustering coefficient of the geo-social network, with the general conclusion of \eqref{constantCluster}, that at high popularity factor the clustering coefficient becomes independent from the node popularity. Here we show numerical results, and evaluate the effects of parameters $\alpha$, $\beta$ and $\delta$.  We evaluate also how the clustering coefficient depends on the node degree. For the simulation results, we sample the popularity of the typical node according to the linear scale, and perform 10 simulations with random topology for each considered $m$ value.

We see that the analytic results fit well the simulation ones. The clustering coefficient decreases with the node popularity, as well as with the node degree, as it is typical in scale-free networks. However, only until the limiting value of $m$ is reached. At this point, however, the node degree itself is not affected by $m$, and we see that all the high popularity nodes have the same degree as well as clustering coefficient. That is, the geographic distance has significant effect on the clustering properties of the network, keeping the high popularity nodes clustered.

Smaller popularity exponent $\beta$ results in higher clustering
coefficients in general, since at small $\beta$ values there are fewer
nodes with very low popularity, and therefore neighbors of a node are more
likely to connect to each other.
The geographic constraints affect the clustering as well through the rank exponent $\alpha$, and the accessibility radius $\delta$. Considering $\delta$, it has no influence at low popularity values, when the social ties are anyway short. However, the clustering coefficient in a network with smaller $\delta$ reaches the limiting value at smaller $m$, and consequently remains to be higher for the high popularity nodes. The rank exponent $\alpha$ has conflicting effects. At low popularity factors, lower $\alpha$, that is, less spatial constraint, allows longer links, which leads to lower clustering. However, the limiting value is reached at a smaller $m$, and as a result, under low $\alpha$ the  clustering coefficient flattens out at a relatively high value.

\section{Discussion of analytic and measurement results}\label{sec:results_analysis}

Let us validate the characteristics of the proposed network model utilizing large scale measurement results from the literature \cite{newman2002assortative,holmeSN14,ahn2007analysis,misloveIMC07,hu2009disassortative,szellSN10,liTBD17,muchnik2013}.

%

Many of the measured large social networks show power-law degree distribution \cite{misloveIMC07,muchnik2013,liTBD17}, supporting out analysis. However, in some cases the decay of the node degrees is faster, \cite{holmeSN14,szellSN10}, showing the need of modeling the limiting effects on link formation.

Measurements on local clustering coefficients agree with the results presented here, showing that the clustering coefficient is rather small and is decreasing with increasing node degree \cite{ahn2007analysis,misloveIMC07,szellSN10,liTBD17}. Measurement results in the high node degree region are however not really reliable due to the small number of the samples.

Since \cite{newman2002assortative} there has been a common agreement that social networks are assortative. This has been confirmed for small size professional networks. Measurement results on on-line social networks, however, often show a transition, and large networks are disassortative \cite{holmeSN14,ahn2007analysis,hu2009disassortative}, a result that complies with our conclusions. Still, many measurements, often considering networks of smaller size, show that the networks are neutral \cite{ahn2007analysis,misloveIMC07,szellSN10,liTBD17}. This shows that further studies are needed on attachment preferences. The introduction of additional attributes, as proposed in \cite{gongIMC12} could be a powerful approach.

\section{Conclusion}\label{sec:conclusion}

In this work we propose a geo-social network model, where social ties develop according to
the individuals' popularity and spatial distance, the population density and the communication range, \textit{i.e.} accessibility radius. We consider power-law distributed popularity, and model the geographically distributed individuals with a homogeneous marked Poisson point process, with their popularity as marks.
We derive analytic models to characterise the emerging networks.

Based on the analytic and numerical results, we conclude that the geography
affects the network structure, with the spatial communication range as the most significant factor.
We show that the geographic limitations allow power-law degree distribution. The power-law
distribution is affected by both individuals' popularity distribution and
also by the rank exponent, but turns out to be independent of population density.
Low popularity nodes do not experience much from this change, they experience degree distribution, neighbor degree distribution and cluster coefficient as typical in non-geographic scale-free networks. However, highly popular nodes are affected by the constraints of the geography, and experience a random lattice like environment, even if the node degree distribution in the network remains power law. If the connection range is strongly limited, the network becomes a random lattice, even under the mutual interest based attachment.

\ifCLASSOPTIONcompsoc                                                                                       
  \section*{Acknowledgments}                                                                                
\else                                                                                                       

  \section*{Acknowledgment}                                                                                 
\fi                                                                                                         
This research was in part supported by the Swedish Research Council.              

\bibliographystyle{IEEEtran}

\bibliography{IEEEabrv,social_tie}

\vfil
\vfil
\begin{IEEEbiography}
[{\includegraphics[width=1in,height=1.25in,clip,keepaspectratio]{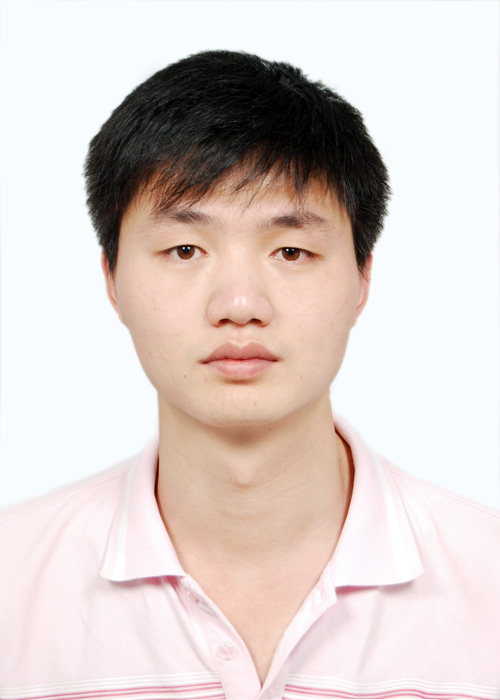}}]
  {Dong Liu} received M.Sc. degree in Information and Communication
  Engineering from Tongji University, Shanghai, China, in 2016. He is
  currently pursuing the Ph.D. degree in the Department of
  Information Science and Engineering at the School of Electrical
  Engineering and Computer Science, KTH Royal Institute of Technology, Stockholm,
  Sweden. His research interests include stochastic network modeling,
  complex networks, optimization and machine learning.
\end{IEEEbiography}
\vfil
\begin{IEEEbiography} [{\includegraphics[width=1in,height=1.25in,clip,keepaspectratio]{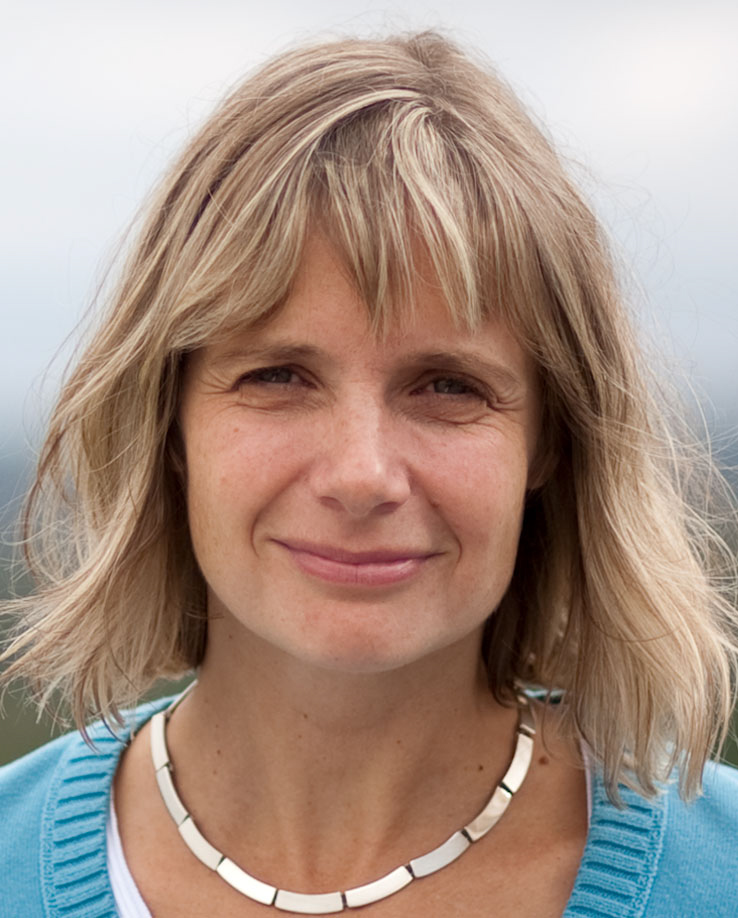}}]
  {Viktoria Fodor}
  is professor at KTH Royal Institute of Technology, Stockholm, Sweden. She received her M.Sc. and Ph.D. degrees in computer engineering from the Budapest University of Technology and Economics in 1992 and 1999, respectively. She worked at the Hungarian Telecommunication Company in 1998 and joined KTH in 1999. She is associate editor at IEEE Transactions on Network and Service Management, and the Transactions on Emerging Telecommunications Technologies. In 2017 she acted as co-chair of IFIP Networking. Her current research interests include network modeling and performance evaluation, protocol design, wireless and multimedia networking.
\end{IEEEbiography}
\vfil
\vspace*{-2\baselineskip}
\begin{IEEEbiography}
[{\includegraphics[width=1in,height=1.25in,clip,keepaspectratio]{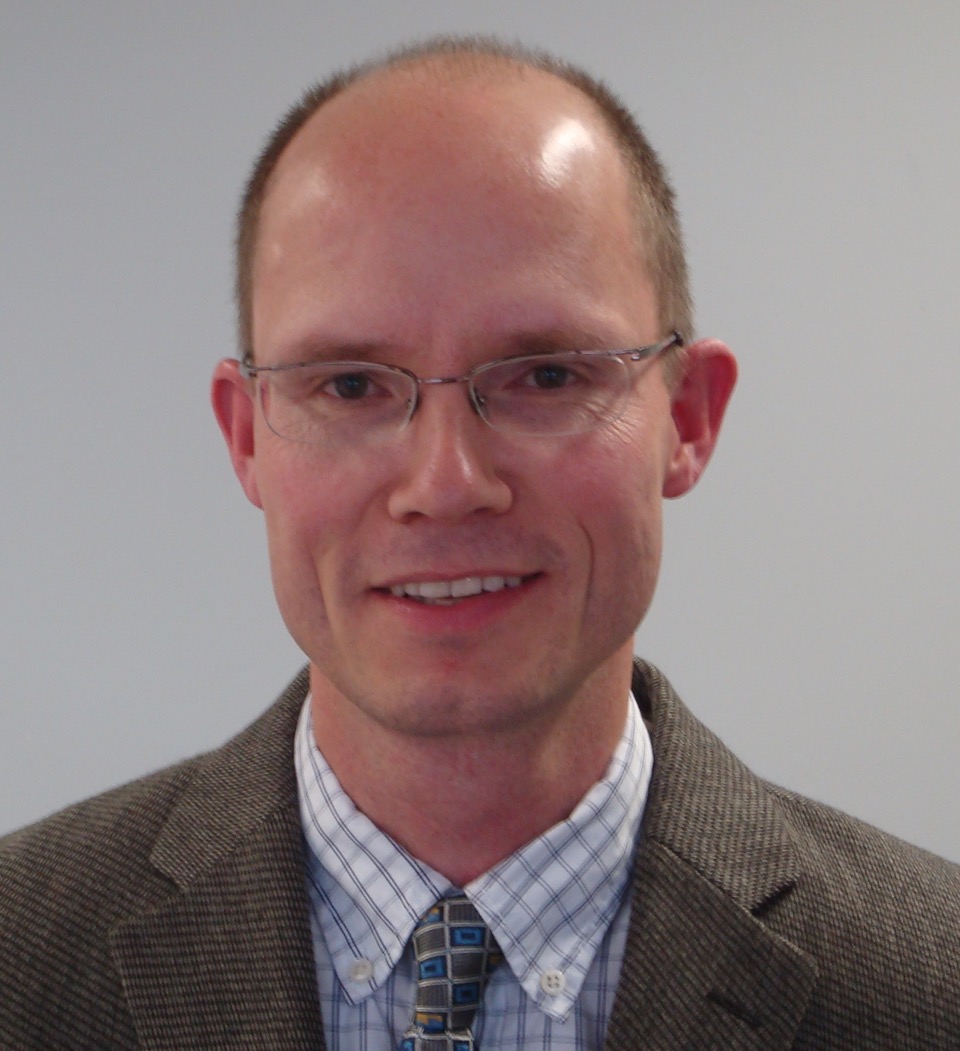}}]{Lars Kildeh{\o}j Rasmussen} (S92, M93, SM01) got his M. Eng. in 1989 from the Technical University of Denmark (Lyngby, Denmark) and his Ph.D. degree from Georgia Institute of Technology (Atlanta, Georgia, USA) in 1993. He is now a Professor in the Department of Information Science and Engineering at the School of Electrical Engineering, and the ACCESS Linnaeus Center, at the KTH Royal Institute of Technology (Stockholm, Sweden).
  He has prior experience from the Institute for Telecommunications
  Research at the University of South Australia (Adelaide, Australia),
  the Center for Wireless Communications at National University of
  Singapore (Singapore), Chalmers University of Technology (Gothenburg,
  Sweden), and University of Pretoria (Pretoria, South Africa).
  He is a co-founder of Cohda Wireless Pty Ltd
  (http://www.cohdawireless.com/); a leading developer of Safe Vehicle
  and Connected Vehicle design solutions.
  He is a Senior Member of the
  IEEE, a member of the IEEE Information Theory Society, Communications
  Society, and Vehicular Technology Society. He served as Chairman for
  the Australian Chapter of the IEEE Information Theory Society
  2004-2005, and has been a board member of the IEEE Sweden Section
  Vehicular Technology, Communications, and Information Theory Joint
  Societies Chapter since 2010. He is an associate editor for IEEE
  Transactions on Wireless Communications, and Elsevier Physical
  Communications, as well as a former associate editor of IEEE
  Transactions on Communications 2002-2013. His research interests include transmission strategies and coding schemes for wireless communications, communications and control, vehicular communication systems, and signal and information processing over networks.
\end{IEEEbiography}

\end{document}